# Wide temperature span of entropy change in first-order metamagnetic MnCo$_{1-x}$Fe$_x$Si

**J H Xu[1], W Y Yang[1], Q H Du[1], Y H Xia[1], X M Liu[2], H L Du[1], J B Yang[1], C S Wang[1], J Z Han[1], S Q Liu[1], Y Zhang[1], Y C Yang[1]**

[1] School of Physics, Peking University, Beijing 100871, People's Republic of China

[2] School of Physics and Engineering, Zhengzhou University, Zhengzhou 450052, People's Republic of China

E-mail: liuxmamara@zzu.edu.cn

**Abstract.** The crystal structure and magnetic properties of MnCo$_x$Fe$_{1-x}$Si ($x$=0-0.5) compounds were investigated. With increasing Fe content, the unit cell changes anisotropically and the magnetic property evolves gradually: Curie temperature decreases continuously, the first-order metamagnetic transition from a low-temperature helical antiferromagnetic state to a high-temperature ferromagnetic state disappears gradually and then a spin-glass-like state and another antiferromagnetic state emerge in the low temperature region. The Curie transition leads to a moderate conventional entropy change. The metamagnetic transition not only yields a larger negative magnetocaloric effect at lower applied fields than in MnCoSi but also produces a very large temperature span (103 K for $\Delta\mu_0 H$=5 T) of $\Delta S(T)$, which results in a large refrigerant capacity. These phenomena were explained in terms of crystal structure change and magnetoelastic coupling mechanism. The low-cost MnCo$_{1-x}$Fe$_x$Si compounds are promising candidates for near room temperature magnetic refrigeration applications because of the large isothermal entropy change and the wide working temperature span.

## 1. Introduction

Recently, room-temperature magnetic refrigeration based on the magnetocaloric effect (MCE) has emerged as a competitive technology to gas compress for it is energy efficient and also environmentally friendly. Up to date, there are several candidate materials for magnetic refrigeration and according to the nature of the magnetic transition, they fall into two categories: first-order magnetic transition (FOMT) materials and second-order magnetic transition (SOMT) materials. The former exhibits the so-called giant magnetocaloric effect (GMCE) usually with a narrow temperature span due to a magneto-structural or a magneto-elastic coupling effect, for example, GdSiGe [1], LaFeSi [2], MnFePSi [3],MnNiGe [4]. But some forms of hysteresis in these systems are detrimental to the efficiency of a cooling cycle. In contrast, the latter－e.g. Gd [5, 6], most of the manganites [6] and numerous intermetallics [5, 6]－shows a less intense but broader $\Delta S(T)$ peak and no obvious hysteresis. However, the refrigerant capacity (RC) (the amount of heat which can be transferred in one thermodynamic cycle, taking account of the width and height of the $\Delta S(T)$ peak) is a more relevant parameter to evaluate the magnetocaloric properties which points out that a large entropy change with a broad temperature span is favored for applications [6, 7].

Apart from a SOMT from a ferromagnetic state to a paramagnetic state at 390K in MnCoSi, there is another first-order metamagnetic transition induced by temperature or field from a low-temperature helical antiferromagnetic (AFM) state to a high-temperature high magnetization state or ferromagnetic (FM) state [8-10]. It was reported that the metamagnetic transition is highly sensitive to field and produces a large negative magnetocaloric effect [8, 10]. In addition, a slight substitution of Ge for Si lowers the metamagnetic transition temperature and the critical field but doesn't improve the magnitude of the peak MCE entropy [10]. While $MnCoSi_{0.88}Ge_{0.12}$ show a large low-field magnetic entropy change and a broad working temperature span [11]. The doping with Fe on the Mn site also lowers the metamagnetic transition temperature as well as the transition field and yields a higher magnetocaloric effect at lower fields than in MnCoSi [12]. However, the influence of the substitution on the Co site over the crystal structure and magnetic properties is rarely reported. In this paper, we report the results from $MnCo_{1-x}Fe_xSi$ compounds. Similar to the case of the substitution of Fe for Mn, the Fe substitution on the Co site also leads to large changes in the magnetic transitions and more importantly, in the magnetocaloric properties.

2. Experimental

Polycrystalline $MnCo_xFe_{1-x}Si$ compounds with $x$=0-0.6 were prepared by arc melting mixtures of pure elements (99.99%) with a 3% excess of Mn with respect to the 1:1:1 stoichiometry to compensate the amount of evaporation. The obtained ingots were sealed under argon atmosphere in silica tubes and then annealed at 673 K, 973 K, 1100 K, 1200 K, respectively, followed by quenching into icy water. X-ray powder diffraction experiments were performed at room temperature using X'pert Pro MPD diffractometer (Cu $K\alpha$ radiation) and the data were processed by Reitveld refinement using X'pert HighScore Plus software. The magnetization measurements were conducted using a conventional physical property measurement system (PPMS-7) and a magnetic property measurement system (MPMS, Quantum Design).

3. Results and discussion

MnCoSi crystallizes in the orthorhombic TiNiSi-type structure (space group *Pnma*) but the isostructural MnFeSi doesn't exist. Our experiments showed that the addition of Fe indeed lowers the structural stability and at last, results in the formation of an impurity. The as-cast ingots with $x \leqslant 0.5$ were identified by XRD to be a single phase with the TiNiSi-type structure and in contrast, a large amount of impurity of the $Mn_5Si_3$-type structure arises in the ones with $x>0.5$. Moreover, the single-phased samples produce the same impurity if annealed above a critical temperature which lowers with increasing iron content. Therefore, in our experiments, the samples were annealed at 973K for $x\leq0.25$ and at 673K for $0.3\leq x\leq0.5$ to preserve the TiNiSi-type structure of the as-cast samples and at the same time, to release the inner strain. Recently, it was found that strain leads to a large amount of second-order magnetic transition material in the quenched MnCoSi compound and thus a reduction in latent heat as well as an apparent decrease in hysteresis [13]. From the XRD data (figure 1) and the obvious widespread hysteresis of the *M-H* loops (the inset of figure 3), we conclude that our samples are strainless and undergo mostly first-order metamagnetic transition.

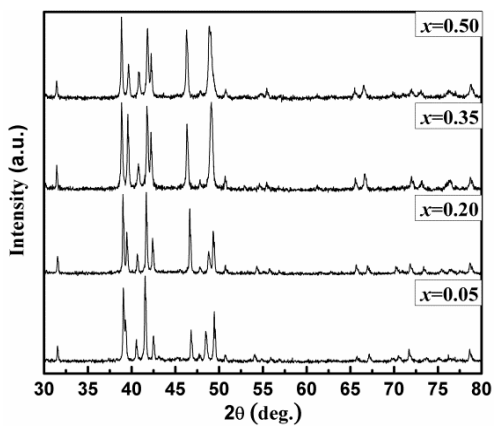

**Figure 1.** Powder XRD patterns measured at room temperature for selected samples ($x$=0.05, 0.20, 0.35, 0.50).

The refined lattice parameter (table 1) shows an evident anisotropic change with increasing Fe content: *a* decreases monotonically while both *b* and *c* change in the opposite tendency. It was reported that the metal-metal bonding via *d* orbitals contributes significantly to the stabilization of the structure [14]. With the increase of Fe concentration, the nearest-neighboring Mn-Mn distance $d_2$ (almost along the $\vec{a}$ axis) decreases from 3.072 Å for $x$=0 to 2.98 Å for $x$=0.5, while the one parallel to the *bc* plane ($d_1$) increases from 3.129Å to 3.295Å. The substitute Fe may enhance the Mn-Mn bond along the $\vec{a}$ axis and thus decrease the length of the $\vec{a}$ lattice vector, which is also happened in Mn(Ni,Fe)Ge [4]. On the other hand, a comparison of the TiNiSi-type structure and the $Mn_5Si_3$-type one may be helpful to understand the change tendency of the cell parameter. Both of the two types of structures can be viewed as being composed of the distorted graphite-like Ni/Mn-Si layers stacked along the $\vec{a}$ axis (for TiNiSi-type structure) or along the $\vec{c}$ axis (for $Mn_5Si_3$-type structure) with Ti or the remaining Mn atoms filling the large holes [15, 16]. One difference between them is that the Ni-Si sheet is puckered in TiNiSi while in the other structure the corresponding sheet is planar, which results in a larger length of the $\vec{a}$ axis of TiNiSi-type structure compared with the length of $\vec{c}$ axis of the $Mn_5Si_3$-type structure [15]. Since both $Mn_5Si_3$ and $Fe_5Si_3$ are isotopic with the structure of $Mn_5Si_3$, the change of the Mn(Co,Fe)Si cell with Fe content and the appearance of the impurity may be understandable.

**Table 1.** Lattice parameters and magnetic transition temperatures $T_c$, $T_t$ and $T_g$ (spin-glass-like transition temperature) for MnCo$_{1-x}$Fe$_x$Si compounds.

| x | a(Å) | b(Å) | c(Å) | V(Å$^3$) | $T_c$ (K) | $T_t$/$T_g$(K) |
|---|---|---|---|---|---|---|
| 0.05 | 5.850 | 3.693 | 6.867 | 148.353 | 377 | 240 |
| 0.10 | 5.839 | 3.696 | 6.874 | 148.376 | 366 | 188 |
| 0.15 | 5.829 | 3.698 | 6.881 | 148.333 | 338 | - |
| 0.20 | 5.826 | 3.700 | 6.887 | 148.460 | 324 | - |
| 0.30 | 5.813 | 3.705 | 6.897 | 148.531 | 314 | - |
| 0.35 | 5.766 | 3.718 | 6.920 | 148.312 | 288 | 13($T_g$) |
| 0.40 | 5.780 | 3.720 | 6.914 | 148.670 | 260 | 21($T_g$) |
| 0.50 | 5.756 | 3.735 | 6.932 | 149.043 | 212 | 17($T_g$),140($T_t$) |

The thermomagnetic curves (*M-T* curves, 5K~350 K in a field of 50 Oe) under zero-field-cooled (ZFC) condition and field-cooled (FC) condition for selected samples were showed in figure 2. The thermomagnetic behavior changes significantly with Fe content: Curie temperature $T_c$ decreases rapidly from 390K (*x*=0) [8] to 212K (*x*=0.5); the first-order metamagnetic transition from the AFM state to the FM state is shifted to lower temperatures and then disappears for 0.15⩽*x*⩽0.35 (i.e. the AFM state vanishes). Moreover, when *x*⩾ 0.35, a spin-glass-like (SGL) state appears in the low temperature region and for *x*=0.5, another AFM state is also shown by the decrease of the magnetization in the ZFC and FC *M-T* curves. The SGL transition was also evidenced by the temperature dependent of ac susceptibility data measured at several frequencies (not shown here). In addition, the absence of the thermal hysteresis between the ZFC and the FC *M-T* curves around $T_c$ for all the samples points to a second-order magnetic transition and the strong shift between them is ascribed to domain wall pinning [17]. On the other hand, the minor thermal hysteresis around $T_t$ indicates a weak first-order magnetic transition.

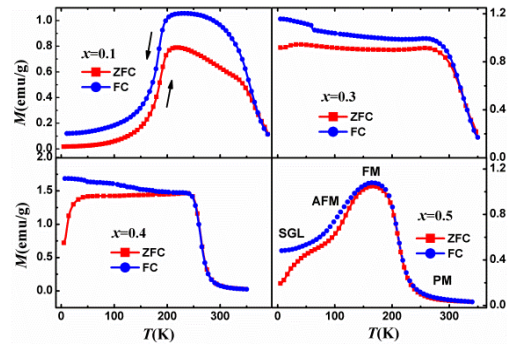

**Figure 2.** Temperature dependence of the magnetization for the representative samples on heating and cooling under a field of 50 Oe.

According to the *ab-initio* calculations for Mn-based orthorhombic magnets in the *Pnma* space group, the nearest-neighboring Mn-Mn distances $d_1$ plays a crucial role in the magnetism in the related compounds [18, 19]. According to the results from high resolution neutron diffraction experiments, the thermal evolution of the helimagnetic state is accompanied by a giant opposite change of the two distances between nearest Mn atoms [20] and the field-induced metamagnetic transition results in an abrupt jump in the lattice parameter and also the Mn-Mn distances [21]. In MnCoSi-based compounds, the AFM state prefers a smaller $d_1$ (i.e. smaller $b$ and $c$) and the FM state a larger $d_1$. The MnCo$_x$Fe$_{1-x}$Si compounds with $x \geqslant 0.15$ have a large room-temperature $d_1$ (due to the increase of $b$ and $c$ with Fe content) so that the distance might not decrease with temperature below a critical value, which may be the cause of the disappearance of the AFM state in the low temperature region.

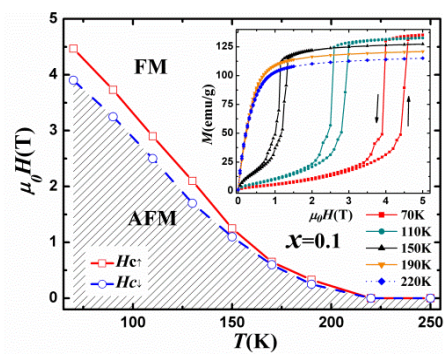

**Figure 3.** Magnetic phase diagram (the main plot) for $x=0.1$ constructed from $M$-$H$ loops and the $M$-$H$ loops (the inset) at various temperatures. $H_{c\downarrow}$ and $H_{c\uparrow}$ indicate the critical field for increasing and decreasing fields.

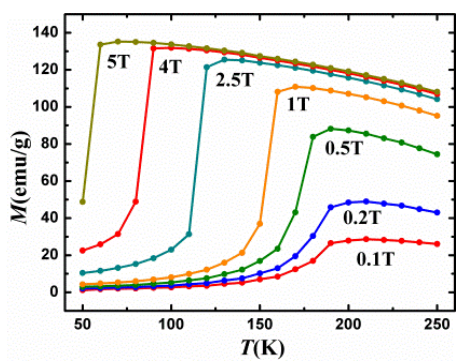

**Figure 4.** Isofield $M$-$T$ curves for $x=0.1$ in various fields (0.1-5T) constructed from isothermal magnetization data.

The main plot of figure 3 shows the temperature dependence of the metamagnetic critical field $H_c$ (obtained from the maxima of the $\partial M/\partial H$ curves) for $x=0.1$ and $H_{c\uparrow}$ and $H_{c\downarrow}$ were defined as the critical field for ascending and descending fields, respectively. The hysteresis width $\Delta H$ of the transition decreases with increasing temperature and vanishes by $T_t$ (~188 K), which can be seen directly from the typical $M$-$H$ loops at some temperatures within 50~250 K (the inset of figure 3). The lowering of the magnitude of the hysteresis at elevated temperatures may be due to the decreases in the magnitude of the lattice parameter jump upon the field-induced transition which leads to a decline in the energy barrier between the two adjacent states [21]. Figure 4 shows the isofield thermomagnetic curves constructed from the isothermal magnetization data. Preserving the highly sensitive metamagnetic transition, the metamagnetic transition can be induced by applying a field of about 5 T at 50 K which is far below $T_t$ and the average $dT_t/dH$ is about -25 K/T which is significantly larger than than of FeRh (-8K/T) [22] and $Mn_3GaC$ (-5K/T) [23].

Figure 5 (a) and (b) present, respectively, the temperature dependences of the isothermal entropy change $\Delta S$ calculated using Maxwell relations around $T_c$ (for $x$=0.35) and around $T_t$ (for $x$=0.1). The $-\Delta S$ around $T_c$ for $x$=0.35 has a mild dependence on temperature with a mediate maximum of 3.5 $Jkg^{-1}K^{-1}$ at a field change of 5 T. On the other hand, a large negative MCE ($\Delta S_{max}$~5.4 $Jkg^{-1}K^{-1}$ for $\Delta\mu_0H$=5 T) were obtained around $T_t$ in $MnCo_{0.9}Fe_{0.1}Si$. Worthy to note is that the first-order metamagnetic transition is softened by Fe substitution, leading to a large magnetocaloric effect with a low field change, e.g. the one with $\Delta\mu_0H$=2 T (figure5 (b)) and singularly broad peaks of entropy change, e.g. the peak widths of 40K and 103K for $\Delta\mu_0H$=2 T and 5T, respectively. Those are significant changes compared with MnCoSi-related materials [10, 12, 21] and especially, the wide temperature span can be comparable to or even larger than that of SOMT materials. In addition, compared with other metamagnetic materials, e.g. FeRh [22], $Mn_3GaC$ [24], the highly field-sensitive metamagnetic transition leads to a lower isothermal $\Delta S(T)$ peak but a larger temperature span [10].

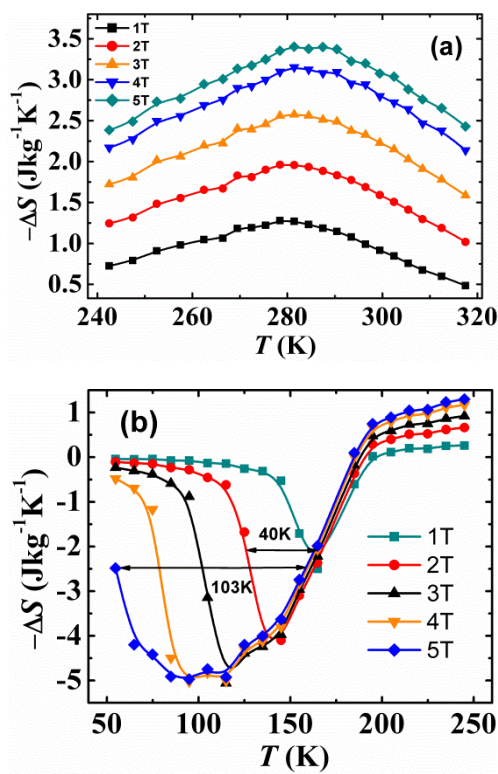

**Figure 5.** Temperature dependence of $\Delta S$ around (a) $T_c$ for $x$=0.35 and (b) $T_t$ for $x$=0.1 at the applied fields of 2 T and 5 T.

Producing an entropy change with a temperature span as large as ~103K for $\Delta\mu_0 H$=5 T is an uncommon event in FOMT materials. This may be due to the magnetic origin of the magnetocrystalline coupling in MnCoSi. The driving force of a first-order magneto-structural transition can be a crystallographic transition, a magnetic transition (when one takes the main role, the other one acts in a cooperative way) or both of them [25, 26]. The high value of $dT_t/dH$ points out that magnetic field is effective in triggering the magnetoelastic transition and a change of the magnetic parameters is the main driving force in MnCoSi. So the spin-lattice coupling mechanism here may be directly associated with the interactions between magnetic atoms and possibly resulted from the correlation between magnetism and the chemical bond. This discussion is consistent with the earlier-stated presume that bonding via $d$ orbits is important for the structural stabilization of TiNiSi-type compounds.

As mentioned above, there is a jump in the lattice parameter upon the field-induced metamagnetic transition and it was also reported that $b$ and $c$ abruptly increase to almost-temperature-independent values while $a$ decreases to a value with a normal thermal dependency. The phenomenon confirms the theoretical result that the magnetic ordering state is highly dependent on the Mn-Mn distance $d_1$ [21]. And this also indicates that there are large opposing magnetoelastic energy and lattice elastic energy in $bc$ plane so that the effect of temperature is almost negligible. Due to the AFM state prefers a smaller $d_1$, the $bc$ plane may be highly compressed in the AFM state and thus the negative thermal expansion of $a$ can be viewed as a compensation to the strongly forced contraction of the $bc$ plane. When a field is applied to the AFM system, a transition to FM state can decrease Zeeman energy and at the same time, lower the lattice elastic energy through the accompanied sudden change in the lattice parameters. In the sense, the change of the lattice can assist the field-induced metamagnetic transition, which may be a part of the reason for the large $dT_t/dH$ and the large temperature span of $\Delta S(T)$. Fe doping might alter the magnetic interactions and the lattice elastic energy and so change the properties of the first-order metamagnetism.

Again, RC is a meaningful parameter to evaluate the MCE. Here, the value of RC is calculated from the product of the maximum ︱$\Delta S$︱ and the full width at half maximum of the entropy change peak which is often called the relative cooling power (RCP) [7].With a wide working temperature span, a large RCP with the values of 168 Jkg$^{-1}$ and 520 Jkg$^{-1}$ for 2 T and 5 T, respectively, were obtained in MnCo$_{0.9}$Fe$_{0.1}$Si compound and they are significantly larger than those of metamagnetic Ni$_{50.3}$Mn$_{35.5}$Sn$_{14.4}$ (40 Jkg$^{-1}$ at 2 T) [27], Ni$_{43}$Mn$_{45}$CuSn$_{11}$ (~28 Jkg$^{-1}$ at 1 T) [28], Mn$_3$GaC (~420 Jkg$^{-1}$ at 5 T) [24]. In addition, the position and width of the entropy change peak are greatly dependent on magnetic field, as shown in figure 5(b), which suggests that MnCo$_{1-x}$Fe$_x$Si compounds would be suitable for magnetic refrigeration applications in wide temperature regions.

**Conclusions**
We have synthesized the TiNiSi-type MnCo$_{1-x}$Fe$_x$Si ($x$=0-0.5) compounds and investigated the crystal structure and magnetic properties. It was found that with increasing iron content, the lattice parameter changes anisotropically. Both $T_c$ and $T_t$ decrease rapidly with $x$ and after the vanishing of the AFM state, a spin-glass-like state and another AFM state appear in the low temperature region. The highly sensitive metamagnetic transition is preserved and also softened, which results in a larger negative MCE at lower fields and a broader MCE peak than in MnCoSi. Moreover, this is the largest working temperature span seen in FOMT materials, which is ascribed to the magnetic origin of the magnetoelastic coupling and the assistance of the lattice elastic energy. The MCE around $T_c$ were also studied for the sample with $x$=0.35 whose Curie temperature is at ambient temperature and found a normal magnetic entropy change of the type of second-order magnetic transition. Because of the high refrigerant performance, the low price and easy fabrication of the compounds, they are promising for magnetic refrigerate applications.

**Acknowledgements**

This work was supported by the National Natural Science Foundation of China (Grant No. 11275013, 50971003 and 51071002) and the National 973 Project (Grant No. 2010CB833104, MOST of China).